\documentclass[prb,twocolumn]{revtex4}

\usepackage{graphicx}
\usepackage{dcolumn}
\usepackage{amsmath}

\def\mg{MgB$_2$}
\def\dos{$N(\epsilon_F)$}

\begin {document}

\title[Short Title]{Pressure dependent thermoelectric power of MgB$_2$ superconductor}

\author{E. S. Choi and W. Kang}
\affiliation{Department of Physics, Ewha Womans University, Seoul
120-750, Korea}
\author{J. Y. Kim, Min-Seok Park, C. U. Jung, Heon-Jung Kim, and Sung-Ik Lee
} \affiliation{National Creative Research Institute Center for
Superconductivity and Department of Physics, Pohang University of
Science and Technology, Pohang 790-784, Korea}

\date{\today}

\begin{abstract}
We have measured temperature dependence of thermoelectric power
(TEP) on MgB$_2$ superconductor under hydrostatic pressure. The
sign and temperature dependence of TEP shows metallic hole
carriers are dominant with activation type contribution at higher
temperature. TEP increases with pressure while $T_c$ decrease with
ratio -0.136 K/kbar. The data are discussed in consideration of
carriers from different bands and anisotropy of compressibility.

\end{abstract}

\maketitle

\section{Introduction}
The mechanism of superconductivity in \mg~ superconductor has been
of a great interest since its discovery.\cite{akimitsu} The
conventional BCS type superconductivity is suggested from
theoretical approaches.\cite{kortus,an} In this model, the light
mass of boron in \mg~ was attributed to have a strong
electron-phonon coupling which enhances $T_c$ of this material. On
the other hand, the hole superconductivity model was also
proposed,\cite{hirsch} where the superconductivity is driven by
undressing of hole carriers.

Experimentally, various experiments have been done on this new
superconductor and some of them could be used as a series of test
for the theoretical models. Apparently, the BCS type
superconductivity is supported by the experimental results of
boron isotope effect,\cite{budko} heat capacity\cite{kremer}, BCS
like superconducting gap\cite{takahashi} and high pressure
measurement.\cite{highp} But the hole superconductivity model also
accounts for the observed results by itself or with some
assumptions.

$T_c$ of BCS superconductor can be expressed as McMillan
formula,\cite{mcmillan} $T_c \sim \omega {\rm exp} (-1/\lambda)$,
where $\omega$ is the phonon frequency and $\lambda$ is the
electron-phonon coupling constant. The pressure effect on $T_c$ is
manifest from $\lambda$ by the relation of $\lambda =
N(\epsilon_F) <I^2>/M<\omega^2>$, where \dos~ is the density of
states at the Fermi level and $<I^2>$ is the electronic matrix
element. By applying pressure, $<\omega^2>$ increases due to
lattice stiffening and \dos~ decreases as the bandwidth increases.
In the hole superconductivity model, $T_c$ was expected to
increase with pressure if the pressure is to decrease in-plain
boron-boron distance. Hall effect measurement under pressure was
suggested to be a crucial test for this model.

Thermoelectric power (TEP) measurements under pressure will be a
good probe to reveal the mechanism of superconductivity and to
investigate transport properties in the normal state. By applying
pressure, band parameters as well as electron-phonon coupling
constant change, which will affect the absolute magnitude and the
temperature dependence of TEP. In the simple metallic band model,
TEP is proportional to \dos~ or inversely proportional to the
Fermi energy.\cite{mcdonald}

We report here the results of TEP measurements under hydrostatic
pressure. The sign of TEP is positive indicating hole carrier is
majority. The temperature dependence is metallic (linear with
temperature) below $\sim$ 120 K, above which negative contribution
begins to appear. The magnitude of TEP increases and $T_c$
decreases with pressure, which cannot be explained by a simple
metallic band model. We tried several models to explain the
temperature dependence of TEP at normal state. And then we adapt
the models to account for the pressure dependence of TEP and
decrease of $T_c$ under pressure.

\section{Experiments}
Polycrystalline sample used in this experiment was prepared at 950
$^{\circ}$C under 3 GPa.\cite{kang}  The superconducting
transition temperature was 38.4 K with transition width of 0.6 K.
The bar shape sample (with dimensions of $4\times0.5\times0.1~{\rm
mm}^3$) was mounted on two resistive heaters. Four gold wires were
attached and two of them were used as thermoelectric potential
leads for TEP measurements and as current feeding leads for
4-probe resistance measurements. Chromel-constantan thermocouples
were used for the temperature gradient measurement. Sample ends
and thermocouple beads were glued to the heater blocks by Stycast
epoxy. 50:50 mixture of Daphne and Kerosene oil was used as a
pressure medium\cite{murata} in the self-clampled BeCu pressure
cell. The pressure dependence of chromel-constantan thermocouples
and gold lead wires were ignored, which are supposed to be very
small. At room temperature, it was estimated that the TEP of
chromel-constantan thermocouple increase by +0.7$\%$, and TEP of
gold decrease by $-0.07 \mu$V/K at 15 kbar.\cite{blatt} As it will
be shown in the following section, the absence of pressure effect
on TEP below $T_c$ justifies our assumption. The detailed
measurement set up will be published elsewhere.

\section{Results and Discussion}
Fig. \ref{fig1} shows the temperature dependence of resistivity
under different pressure. The inset shows the low temperature
expansion around $T_c$. The normal state resistivity and $T_c$
decrease with pressure. $T_c$ was defined as a temperature where
the resistance reduces to one-half of the normal state resistance
of just above $T_c$. The pressure dependence of conductivity
change ratio ($\Delta \sigma/\sigma (1 {\rm bar})$) and $T_c$ was
shown in Fig. \ref{fig2}. $T_c$ decrease with rate $-$0.136 $\pm$
0.004 K/kbar whose value is in well agreement with previous
reports.\cite{highp}

Fig. \ref{fig3} shows the temperature dependence of TEP at ambient
pressure and under pressures of 14.8, 11.3, 9.2, 6.3 and 3.2 kbar
and the inset shows the superconducting transitions. For one
pressure of about 6.5 kbar, the temperature was increased from
room temperature to 370 K, which is also shown in the same figure.
The overall temperature dependence and the magnitude of TEP are
very similar with previous reports,\cite{lorentz} in which ambient
pressure TEP was measured on the sample prepared without high
pressure sintering. TEP is positive in the whole range of
temperature and pressure, which indicates the majority carrier is
hole as also evidenced by Hall effect measurement.\cite{kang} The
overall temperature dependence of TEP is universal regardless of
pressure, i.e., linear $T-$dependence between $T_c$ and $\sim$ 120
K and gradual deviation from linear dependence at higher
temperature. TEP increases with same curvature and same behavior
extends to at least 370 K, the highest temperature in this
measurement.

Superconducting transition is also clearly seen in the TEP
measurements. Below the transition temperature, small offset of
TEP was observed, $\sim$ $-$0.1 $\mu$V/K for 1 bar data and $\sim$
$-$0.2 $\mu$V/K for high pressure data. We believe that they are
due to bad calibrations of gold lead wires in this temperature
range where the temperature dependence of gold wire changes
rapidly due to phonon drag. Nevertheless, the difference of
magnitude of TEP below $T_c$ is within the error bars of our
measurement setup ($\sim$ 50 nV/K) and pressure independent except
1 bar data.

As for the pressure dependence of TEP, both the room temperature
value and the slope of linear $T-$dependence increase with
pressure. In our measurement set up, the pressure was initially
increased from the ambient pressure to the highest value and the
temperature dependence was measured. And then the pressure was
released by few kbars followed by temperature dependence
measurements. Fig. \ref{fig4} shows the pressure dependence of TEP
at room temperature and the slope of linear $T-$ dependence of
TEP. The pressure dependence of TEP at room temperature was
measured during the initial pressurizing and the following
successive temperature dependence measurements. TEP increases
almost linearly by applying pressure with rate
$+4.35(\pm0.09)\times10^{-2}\mu$V/kbar at room temperature.

First, let us look into the temperature dependence of TEP. Due to
the apparent linear $T-$ dependence at the intermediate
temperature region, we are tempted to account for the behavior by
simple metallic diffusion TEP. In the simple metallic one band
model, TEP can be expressed as,\cite{mcdonald,blatt}

\begin{equation}
\label{mott} S =\frac{\pi^2 k_B^2}{3e}T(\frac{n'}{n}+
\frac{\tau'}{\tau})\arrowvert_{\epsilon_{\rm F}}=\frac{\pi^2
k_B^2}{3e}T(N(\epsilon_F)+
\frac{\tau'}{\tau}\arrowvert_{\epsilon_{\rm F}})\end{equation}
where $n$ is the carrier density and $\tau$ is the relaxation
time. If we assume the relaxation time is energy independent, the
linear $T-$dependence can be explained by above formula. According
to the band structure calculations of \mg,\cite{kortus,an} two
dimensional $\sigma$ bands and three dimensional $\pi$ bands,
which originate from the boron $p_{x,y}$ and $p_z$ orbitals
respectively, were shown. The $\sigma$ bands are partially filled
with small dispersion along the $\Gamma$-A line, which can produce
large (large \dos~ from small energy dispersion) and positive
(hole carriers) TEP signal. If we apply Eq. \ref{mott} to derive
\dos~ from the observed linear $T-$ dependence, \dos~ is estimated
to be about 1.42 states/eV-carriers at 1 bar and increase to 1.67
states/eV-carriers at 14.8 kbar. According to An and
Pickett,\cite{an} the density of states of the two dimensional
$\sigma$ band is $N(\epsilon_F)_{\sigma}$=0.25 states/eV-cell and
the total number of holes for that band is 0.13/cell, which
corresponds to 0.25/0.13=1.92 states/eV-$\sigma$ carriers. The
discrepancy of this value with the observed value is not
surprising, because Eq. \ref{mott} is valid for single band and it
should be modified when more than one band transport is involved.

When more than one band contributes TEP, total TEP is expressed
as,\cite{mcdonald}

\begin{equation}
\label{Stot} S_{\rm tot} =
\frac{\sigma_{1}S_{1}+\sigma_{2}S_{2}}{\sigma_{1}+\sigma_{2}}
\end{equation}
where $\sigma_1 (\sigma_2)$ and $S_1 (S_2)$ are conductivity and
TEP from band 1 (band 2). If we have metallic hole band (suppose
band 1) and metallic electron band (suppose band 2), and if we
assume all metallic bands have similar temperature dependence of
conductivity, the total TEP will be expressed as $S_{\rm tot}=A_1
T - A_2 T=(A_1-A_2)T$. A$_1$(A$_2$) is absolute magnitude of slope
of linear $T-$dependence of TEP when TEP of band 1 (band 2) is
probed exclusively. Therefore, TEP resulting from mixing of two
metallic bands will show linear $T-$ dependence again but its sign
and magnitude depends on density of states of each bands. $S_{\rm
tot}$ will be positive if $N(\epsilon_F)_{\rm band 1} >
N(\epsilon_F)_{\rm band 2}$ and negative in opposite case.

 In the band structure calculation, an antibonding $\pi$ band from
boron $p_z$ orbitals is supposed to be electron like, which has
larger dispersion than the $\sigma$ bands. The anitibonding $\pi$
band will give smaller slope of the linear $T-$ dependence with
opposite polarity than the value obtained exclusively from the
$\sigma$ band.

But when we look into the higher temperature region, which
deviates from linear $T-$ dependence, the above simple arguments
can not explain such behaviors. The temperature dependence of TEP
from the $\sigma$ and the antibonding $\pi$ bands give same linear
$T-$ dependence, whereas the deviation becomes greater as the
temperature increases.

The deviation of a linear $T-$ dependence can be originate from
the electron-phonon interaction which was not considered in Eq.
\ref{mott}. The phonon drag is the first candidate, which usually
gives peak structure in the temperature dependence between
$\theta_D$/10 and $\theta_D$/5, where $\theta_D$ is the Debye
temperature. For \mg~ superconductor, $\theta_D$ is estimated
about 900 K.\cite{wang} But in our results, there seems to be no
such effects in the measured temperature range.

The phonon drag effect usually smeared out in the disordered
metal, where the mass enhancement effect by electron-phonon
interaction emerges in TEP instead. This effect was considered to
account for the temperature dependence of high $T_c$
superconductors.\cite{kaiser} TEP with electron-phonon enhancement
can be expressed as, $S_{\rm el-ph} =S_0 (1+\lambda(T))$, where
$S_0$ denotes the bare TEP without enhancement and $\lambda(T)$ is
the electron-phonon coupling constant. the bare TEP behavior is
recovered at high tempertature, since $\lambda(T)$ goes to zero as
phonon-phonon interaction dominates. When considering the fact
that the TEP shows similar behavior up to 370 K, this effect seems
to be less likely in \mg~ superconductor.

Considering the two band model shown in Eq. \ref{Stot}, we can
separate high temperature contribution from the total TEP. If we
assign band 1 as a metallic TEP showing linear $T-$ dependence
(which comes from hole band and/or electron band), the additional
negative contribution (from band 2) can be extracted by
subtracting the linear temperature dependence from the observed
TEP. We plot $\Delta S (=S(T)-A T)$ versus temperature in Fig.
\ref{fig5}, where $A$ is the slope of linear $T-$ dependence of
TEP between $T_c$ and $\sim$ 120 K.

Two notable features are shown in Fig. \ref{fig5}. The magnitude
of $\Delta S$ increases with temperature and pressure, and the
deviation begins to occur above same temperature regardless of
pressure.

The temperature dependence of $\Delta S$ can be explained by
assuming a existence of thermally activated carriers. In the
situation where two types of charge carriers are involved; one is
metallic and the other should overcome a energy gap $E_g$ to
contribute transport, the total TEP can be expressed as,

\begin{eqnarray*}
\label{deltaS} S_{\rm tot}=\frac{\sigma_{1}A T+{\rm
exp}(-Eg/2k_BT)(B+C E_g/2k_BT)}{\sigma_{1}+{\rm exp}(-Eg/2k_BT)}
\end{eqnarray*}
\begin{eqnarray}
\label{deltaS} \approx AT+\frac{{\rm exp}(-Eg/2k_BT)(B+C
E_g/2k_BT)}{\sigma_1}\end{eqnarray}

The approximation is valid when the metallic conductivity is
dominant compared to the activation type transport. The fitting
was done by fixing $\sigma_1$ as $T^n (n=0, -1, -2)$ and
liberating all other parameters. The best fitting curves are shown
in the inset of Fig. \ref{fig5}, where the data were shifted by
$-0.25 \mu$V/K from the lower pressure data one by one. The
fitting curves are obtained for temperature independent
conductivity ($n$=0)\cite{fitting} and the fitted parameters are
listed in Table. \ref{table1}.

$E_g$ is more or less pressure insensitive  while the magnitudes
of $B$ and $C$ increases with opposite sign. The increasing
negative contribution shown in Fig. \ref{fig5} seems to come from
the increase of fitting parameter $B$. To derive physical
properties from the fitting, we speculate the TEP of band 2
follows the TEP of intrinsic semiconductor with band gap $E_g$.
TEP of intrinsic semiconductor can be written as

\begin{equation}
\label{semicon}S_{\rm
semicon.}=-\frac{k_B}{e}[\frac{c-1}{c+1}\frac{E_g}{2 k_BT} +
\frac{3}{4} {\rm ln} \frac{m_e}{m_h}]
\end{equation}
where $c$ is the ratio of the electron to hole mobility
($\mu_e/\mu_h$). Usually when a semiconducting band is dominant in
transport, TEP shows 1/$T$ dependence and its sign exclusively
depends on the mobility ratio. But when other competing bands are
metallic bands, the contribution of TEP of semiconducting band
decrease upon cooling due to the conductivity weighting and its
sign depends on both mobility and effective mass ratio. Comparing
Eq. \ref{deltaS} and Eq. \ref{semicon}, we listed the mobility and
effective mass ratio in Table. \ref{table1}.

The origin of semiconducting hole carriers is not clear yet. One
of the possibilities is that they originate from the lower band of
the $\sigma$ bands when the $\sigma$ bands are splitted into two
sub-bands due to electron-phonon interaction. It is believed that
the $E_{2g}$ mode (in-plane boron-boron displacement) of \mg~
splits the $\sigma$ band,\cite{an} and the lower band is pushed
down to the Fermi level. If the electron-phonon interaction is so
strong, the splitting may push the lower band below the Fermi
level, which results in a semiconducting hole band. Usually the
$\sigma$ bands are believed to take an important role in the
superconductivity of \mg~ superconductor. Therefore, if our
speculation is correct, one should consider reduced density of
states and stronger electron-phonon coupling constants in
theoretical estimation of $T_c$ of this material.

Now, we turn our attention to the pressure effect on normal state
TEP and $T_c$. The increase of slope with pressure seems to be in
apparent contradiction with McMillan formula because \dos~
increases with pressure but $T_c$ decreases. This discrepancy is
due to the fact that the measured \dos~ from TEP contains
contribution from all available bands as mentioned above to
explain the magnitude of slope. If the pressure effect is to
decrease inter-plane distance more significantly than the in-plane
boron-boron distance, the increase of bandwidth of electron band
(assumed to be antibonding $\pi$ band) is more dominant than that
of the $\sigma$ band (in other words, d$N(\epsilon_F)_\pi$/d$P
>$ d$N(\epsilon_F)_\sigma$/d$P$). Consequently, even when the total
density of states decrease (as indirectly evidenced by the
decrease of resistivity with pressure), the measured slope of
linear $T-$ dependence of TEP increases. The compressibility along
the $c$-axis was found to be larger by 64 $\%$ than along the
$a$-axis,\cite{jorgensen} which support this idea.

TEP for Al doped \mg~ was measured by Lorentz et
al.,\cite{lorentz} and they also found an increase of slope and a
decrease of $T_c$ upon electron doping. They attributed the
decrease of $T_c$ to increase of the Fermi energy by doping, since
the Fermi energy is close to an edge of decreasing density of
state curves. But TEP results seems to be contradictory since TEP
decreases for the increased Fermi energy in a simple one band
metal. This inconsistency will be also explained by considering
the role of electron band upon doping.

If we consider the pressure effect on the negative contribution at
high temperature region (above $\sim$ 120 K), the speculation of a
semiconducting hole band should be examined. The pressure will
decrease electron-phonon coupling by lattice stiffening, therefore
the splitting of $\sigma$ band should be suppressed. Consequently,
we expect the semiconducting energy gap ($E_g$) will reduce with
pressure, which is not clearly seen in our measurements (see
Table. \ref{table1}). The pressure effect seems to be rather
dominant in the mobility and the effective mass term. It might be
necessary to consider more complex situation like inter-band
charge transfer to explain the observed TEP at high temperature
region.

Within our model, we expect the anisotropic TEP measuremetns
and/or the measurements under uniaxial stress will be helpful to
understand the current observation. The sign of TEP and the
temperature dependence of TEP will be quite different between
in-plane and inter-plane direction. The in-plane TEP will be more
or less same with polycrystalline sample data if our model is
correct and the inter-plane TEP can be negative and linear at
whole temperature range. And if a single crystal sample is
compressed only along the $c$-axis while liberating expansion
along the $ab$-plane, or vice versa, the behaviors of normal state
TEP and $T_c$ will show different behaviors depending on the
direction of compression.

\section{Conclusion}
The temperature dependence of TEP under hydrostatic pressure was
measured on \mg~ superconductor. The results show that metallic
hole carriers (presumably from the $\sigma$ bands) are important
to explain temperature dependence of TEP. But the pressure
dependence of $T_c$ and TEP needs contribution of carriers from
other bands, i.e., metallic electron band and semiconducting hole
band. The increase of TEP and the decrease of $T_c$ can be
explained by the two metallic bands (hole and electron band). The
density of states of both bands will decrease with pressure but
the electron band will give larger effect due to the anisotropic
compressibility. The deviation from the metallic TEP can be
explained by the semiconducing hole band, whose contribution is
only observable at higher temperature.

\begin{acknowledgments}
This work was supported by the grant for the promotion of
scientific research in women's universities (00-B-WB-06-A-04) and
by the Brain Korea 21 Project in 2000. The pressure setup used in
this study was developed with the support from the Ewha Research
Fund of the year 2000. Works done at POSTECH was supported by the
Ministry of Science and Technology of Korea through the Creative
Research Initiative Program.
\end{acknowledgments}

\newpage


\newpage
\begin{table}
\caption{Fitting parameters obtained from Eq. \ref{deltaS} to fit
curves of Fig. \ref{fig5}. See text for the meaning of
parameters.} \label{table1}
\begin{tabular}{cccccc}

$P$(kbar)&$E_g$(K)&$B$($\mu$V/K)&$C$($\mu$V/K)&$\mu_e/\mu_h$&$m_e/m_h$\\
\colrule
\\ambient&1300&-46.9&8.7&0.817&2.05\\
3.2 &1280&-49.4&8.9&0.814&2.13\\6.3&1230&-48.8&9.6&0.801&2.11
\\ 9.2 &1270&-52.2&9.6&0.800&2.23
\\ 11.3&1270&-52.8&9.9&0.795&2.24
\\ 14.8&1270&-54.8&10.3&0.788&2.32\\
\end{tabular}
\end{table}

\begin{figure}[bp]
\caption{Temperature dependence of resistivity of \mg~ under
hydrostatic pressure. The inset shows expansion around $T_c$. The
pressure shown here is 1bar, 3.2, 5.0, 5.7, 6.9, 8.8, 10.4, 12.2,
13.9 and 14.6 kbar from top to the bottom in the larger figure and
from right to the left in the inset.}\label{fig1}
\end{figure}
\begin{figure}[bp]
\caption{(a) Change of conductivity ($\Delta \sigma/\sigma (1 {\rm
bar})$) at room temperature as a function of pressure (b) $T_c$
versus pressure measured from the resistivity
measurements}\label{fig2}
\end{figure}
\begin{figure}[bp]
\caption{Temperature dependence of TEP of \mg~ under hydrostatic
pressure. TEP increases as pressure varies as 1 bar, 3.2, 6.3,
9.2, 11.3 and 14.8 kbar. The data for temperature above 300 K was
measure at 6.5 kbar. The definition of $\Delta S(T)$ is also
shown. The inset shows low temperature expansion around
$T_c$}\label{fig3}
\end{figure}
\begin{figure}[bp]
\caption{Room temperature TEP and the slope of linear
$T$-dependence of TEP as a function of pressure. Room temperature
TEP was measured during initial pressurization ($\triangle$) and
after temperature dependence measurements ($\circ$).} \label{fig4}
\end{figure}
\begin{figure}[bp]
\caption{Temperature dependence of $\Delta S(T)(=S(T)-AT)$ at
different pressure. The inset shows fitting curves obtained from
Eq. \ref{deltaS}. The data in the inset were shifted by downward
to show temperature dependence explicitly.} \label{fig5}
\end{figure}

\begin{thebibliography}{999}
\bibitem{fitting} One should be careful with the fitting by the two-band model, since
three fitting parameters combined with polymomials and exponents
can generate various curves. Therefore, we determined the best
fitting results whose parameters have systematic pressure
dependence.
\end{thebibliography}
\end {document}
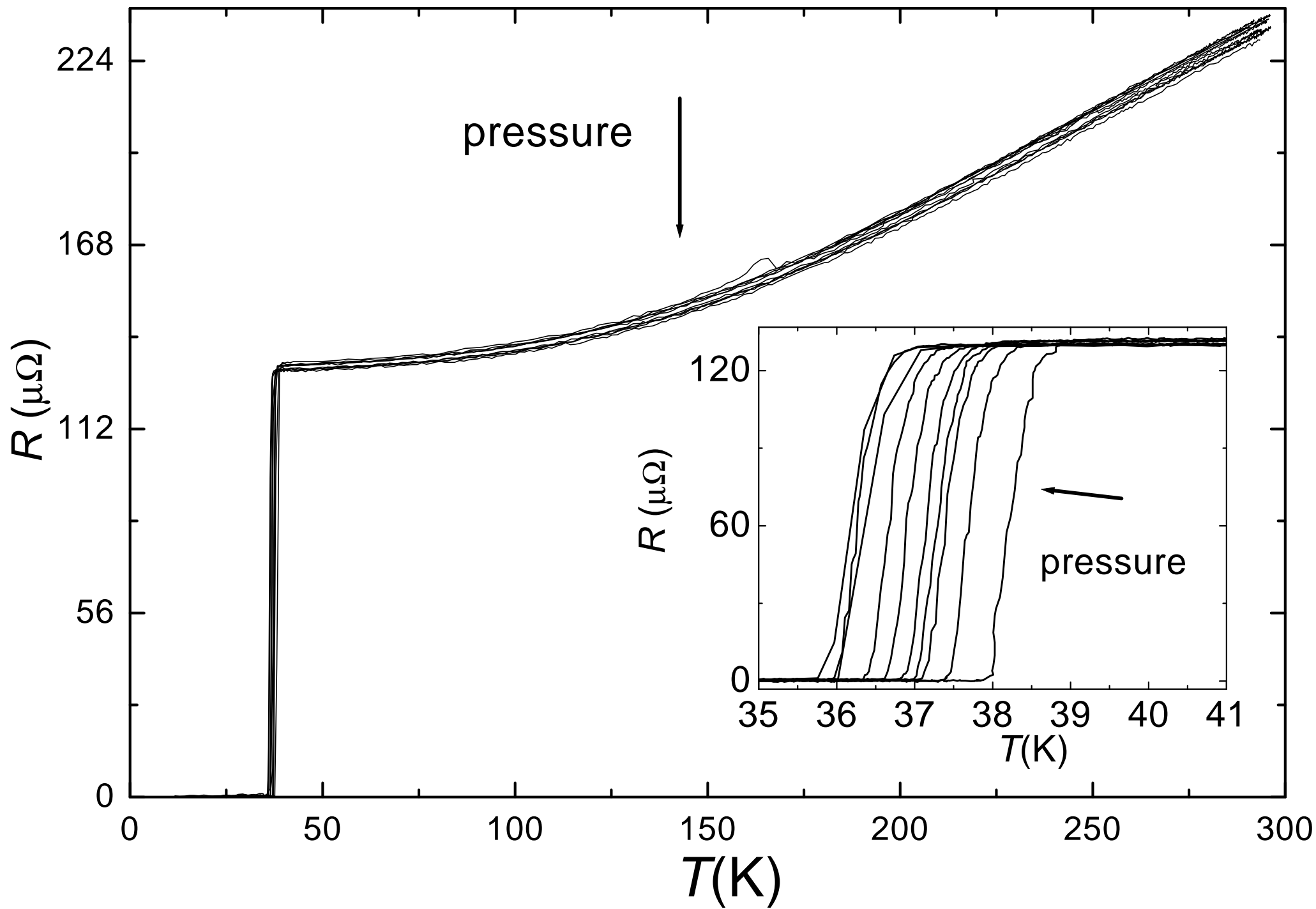
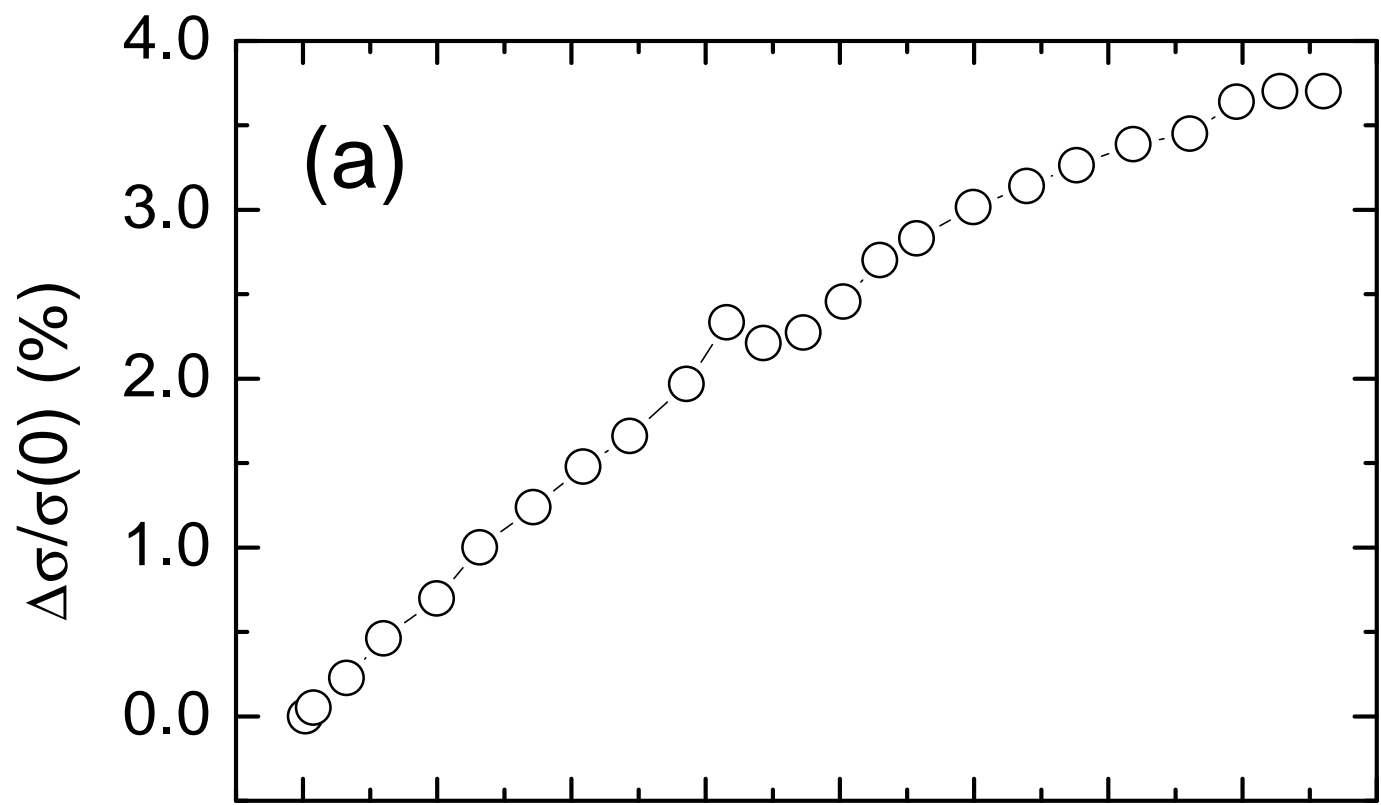
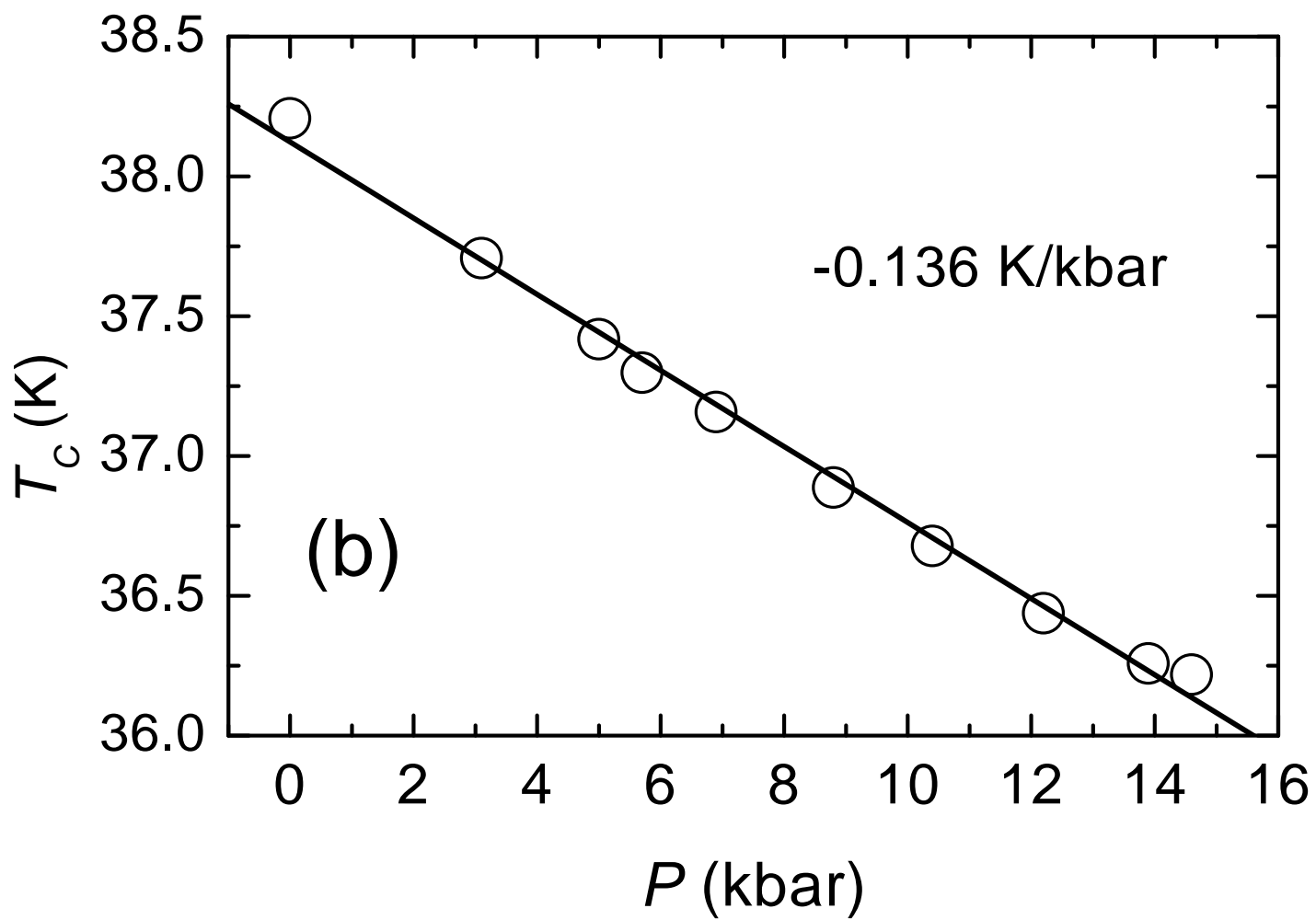
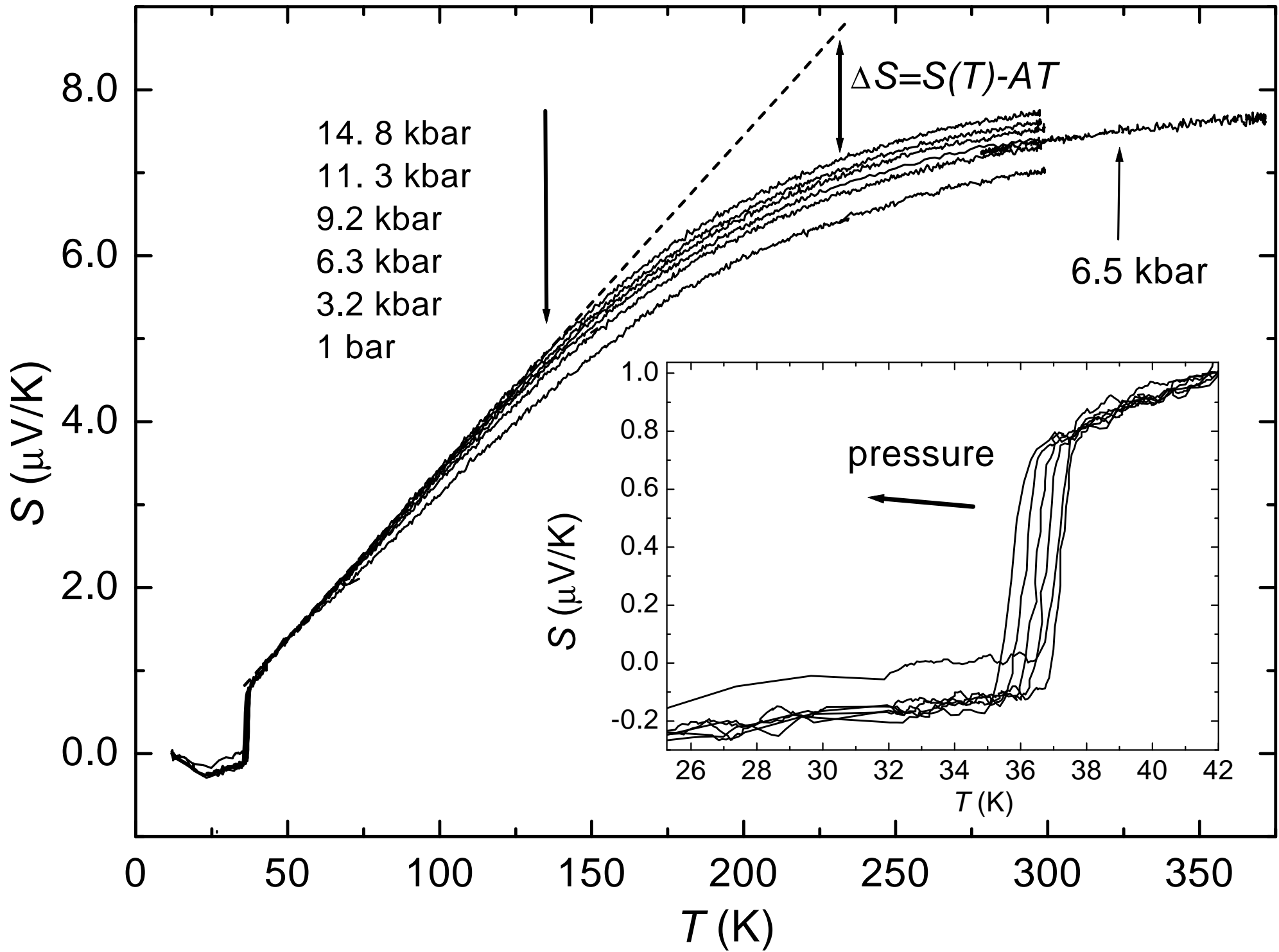
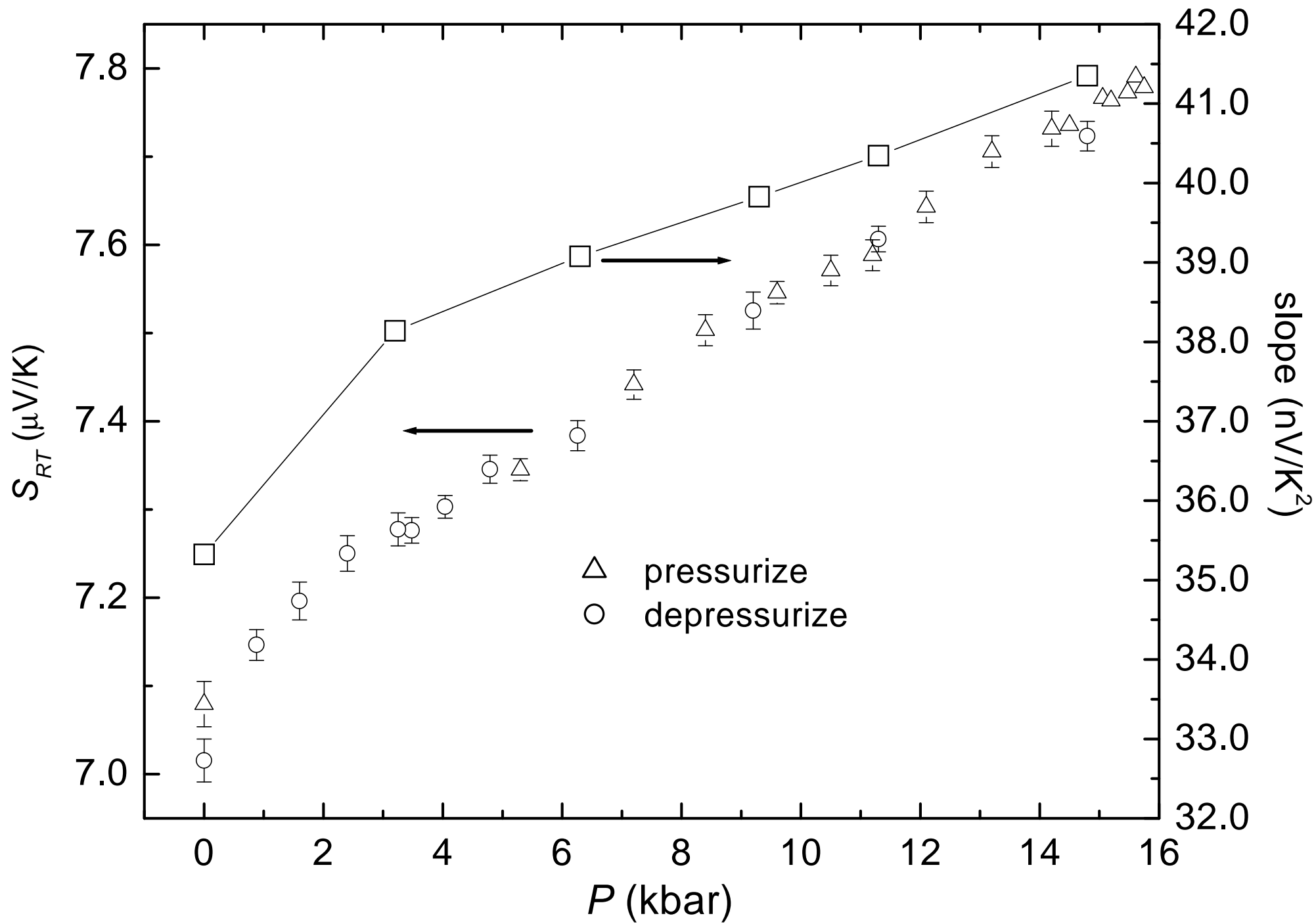
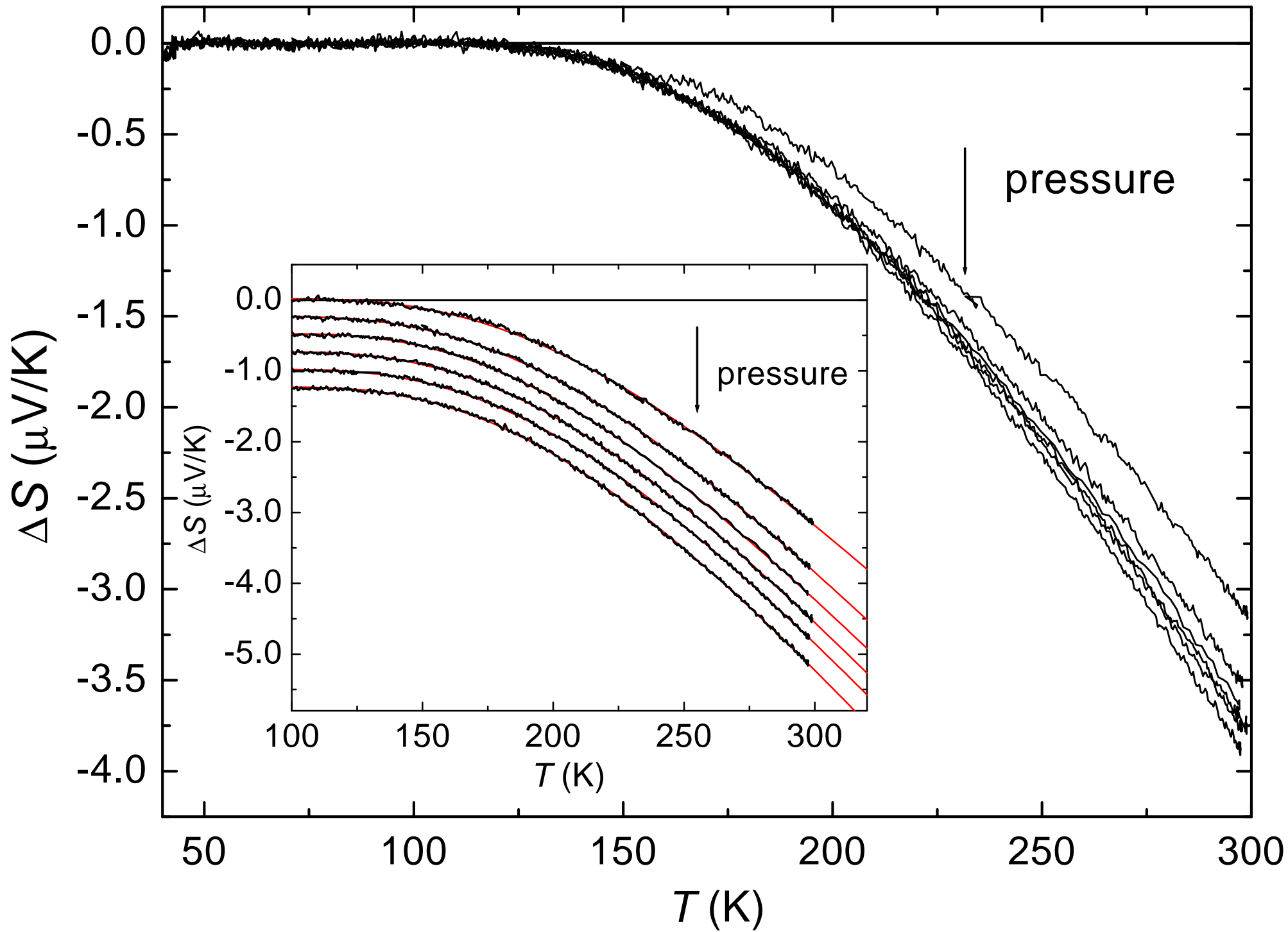
